\documentstyle[12pt]{article}
\title{\bf
	     STOCHASTIC RESONANCE IN ISING SYSTEMS}
\author{
{\bf   Zolt\'an N\'eda}\thanks
{Permanent address: {\it
      Babe\c {s}-Bolyai University , Dept. of  Physics,
      str. Kog\u {a}lniceanu 1, RO-3400 Cluj-Napoca, Romania }  } \\
{\small\it
       University of Bergen, Section for Theoretical Physics }\\
{\small\it
	 All\'egaten 55, N-5007 Bergen, Norway}  }

\date{}

\begin{document}

\maketitle

\begin{center}

Abstract\\
\end{center}

We study by Monte Carlo techniques the evolution of finite two-dimensional
Ising systems in oscillating magnetic fields. The phenomenon of stochastic
resonance is observed. The characteristic peak obtained for the
correlation function between the external field and magnetization,
versus the temperature of the system, is studied for
various external fields and lattice sizes.

\vspace{0.5in}
PACS number(s): 05.40+j; 75.10+h

\vfill \eject

A periodically modulated bistable system in the presence of noise exhibits
the phenomenon of stochastic resonance (SR) [1-3]. Plotting the
correlation $\sigma$ between the modulation signal and the response of the
system, versus the noise intensity we usually obtain a strong peak. This
characteristic picture is accounted in the literature as the phenomenon of SR.
A brief review of the theoretical aspects is given in [4]. Experimental
evidence for the SR phenomenon was found in analog simulations
with proper electronic circuits [3,5,6], laser systems operating in
multistable conditions [7], electron paramagnetic resonance [8,9]
and in a free standing magnetoelastic ribbon [10].

In the present paper we report on the possibility of obtaining the
SR in a bidimensional Ising system.
The phenomenon of SR was already studied in globally coupled two-state
systems [11,12]. Due to the fact that the spins in the Ising model can
be considered as coupled bistable elements, the proposed system is
special case of the problem treated in [11].
In contrast with all earlier
works, we do not consider any external stochastic forces, just the
thermal fluctuations in the system.

Considering an Ising system on a finite square lattice at zero thermodynamic
temperature, the free-energy curve versus the magnetization will have a
double well form [13]. An external oscillating magnetic field will
modulate the two minimums in antiphase. The effect of a positive temperature
in this system can be considered as a stochastic driving force, and the
magnetization in function of the time ($m(t)=\sum_i S_i^z$) as the response
function of the
system. In this way all the necessary conditions are
satisfied for the SR phenomenon. Due to the fact that the noise intensity
(thermal fluctuations) is temperature dependent, the characteristic peak must
be observed for a $T_r$ resonance temperature in the plot of the correlation
($\sigma$) versus the temperature ($T$).

It worth mentioning that recently the Ising system in oscillating fields
was considered by computer simulations, studying the hysteretic response
of the system [14].

We studied the SR in the proposed system by a Monte Carlo method using the
well-known
Metropolis algorithm [15]. The Hamiltonian of the problem is:
\begin{equation}
H=-J\sum_{i,j}S_i^zS_j^z+\mu_B B(t) \sum_i S_i^z,
\end{equation}
where the sum is refering to all nearest-neighbours, $S_i^z=\pm 1$,
$\mu_B$ is the Bohr magneton and $B(t)$ the external magnetic field.
We will consider $B(t)$ in a harmonic form:
\begin{equation}
B(t)=A\cdot  \sin{(\frac{2 \pi}{P} t)}
\end{equation}
The time-scale was chosen in a convenient form, seting the unit-time interval
equal to the average characteristic time ($\tau$) necessary for the flip of
a spin. We have taken this $\tau$ time interval as a constant, and thus
independent
of the temperature. Although this assumption is just a working hypothesis,
we expect usefull qualitative results.

Our Monte Carlo (MC) simulations were performed on square lattices with
$N\times N$
spins, considering the value of $N$ up to $200$. One MC step is defined as
$N\times N$
trials of changing spin orientations and corresponds to a $\tau$ time-interval.
(The period ($P$) of the oscillating magnetic field is also given in this
$\tau$
units.) The amplitude ($A$) of the magnetic field is considered already
multiplied with $\frac{\mu_B}{k}$,
and thus it has the dimensionality of the temperature ($k$ is the Boltzmann
constant).
The temperature is given in arbitrary units. The critical temperature of the
infinite system
\begin{equation}
T_c=2.2692...\frac{J}{k},
\end{equation}
is always considered as $100$ units.
Starting the system from a completely random configuration, to approach the
dynamic
thermodynamic equillibrum we considered $5000$ MC steps. The correlation
function
between the driving field and the magnetic response of the system
\begin{equation}
\sigma=<B(t) \cdot m(t)>=\frac{1}{n} \cdot \sum_{i=1}^{n} B(t_i)\cdot m(t_i),
\end{equation}
was studied during $n=5000$ extra iterations. (The averaging in the
expression of $\sigma$ is in function of time, and $t_i=\tau \cdot i$)
The correlation (4) was studied in function of:
\begin{itemize}
\item the temperature ($T$)
\item lattice size ($N$)
\item amplitude of the magnetic field ($A$)
\item period of the magnetic field ($P$)
\end{itemize}

Our results are summarized in Fig.1-4. As we expected for a fixed lattice size
and a given oscillating magnetic field the curve $\sigma$ versus $T$ exhibits
the characteristic peak of the SR. Considering $A=10$,  $P=50$ and $N=5$
a generic result is ploted in Fig.1.
{}From Fig.2 we conclude that the location of the peak is not significantly
influenced by
the amplitude of the magnetic field. As expected the amplitude influenced
only the shape of the peak, and its height is stongly increasing with the
amplitude. (Our results would suggest an exponential type variation.)
For small lattices ($N<10$) this resonance temperature ($T_r^N$) exhibits a
strong
dependence of the lattice size, and converge to a $T_r$ limit value for
big lattices ($N>100$):
\begin{equation}
T_r=lim_{N\rightarrow \infty} T_r^N
\end{equation}
In Fig.3 we present the results for three choices of $P$.
One can immediately observe that the $T_r$ resonance temperature is
dependent on the period of the magnetic field. As we illustrated in Fig.4
this dependence can be approximated with the curve:
\begin{equation}
T_r=T_c+557.264\cdot P^{-0.6335}
\end{equation}
In the limit of high frequencies the resonance temperature is tending to
infinity and in the limit of small ones, $T_r$ is tending to $T_c$. Due to the
fact that the real experimental conditions work in the very low frequency
limit (the periods in Fig.4 are given in units of $\tau$!), one would expect
this phenomenon detected at $T_c$.

Due to the sensibility of $T_r$ on the size of the system (Fig.3 for small $N$)
an
interesting subject would be the study of this effect on materials containing
small
magnetic domains.

The theory of SR has been previously developed in the context of classical
statistical physics using linear response theory and the
fluctuation-dissipation relations [16,17]. In this sense,
by calculating the susceptibility of our model in the absence
of the periodic magnetic field and studying the response of the system
with the linear response theory one could perform analytical
studies of the problem. We consider that such studies would be
important both for statistical physics and for the phenomenon of SR.

{}From computational point of view, a much more
detalied study on the phenomenon, and the consideration of the real $3D$ case
would also be of interest.

In conclusion, in this letter we studied by MC techniques the phenomenon of SR
in finite Ising systems considered in a periodic magnetic field. The thermal
fluctuations were considered as a stochastic force, and the resonance was
detected for a $T_r$ resonance temperature. The dependence of $T_r$ against the
characteristics of the magnetic field and the lattice size were investigated.

The MC simulations were performed on the computers of the Theoretical Physics
Department (SENTEF) from the University of Bergen (Norway). I am greatfull
to L. Csernai for his continuous help.

\newpage

\section*{References}
\begin{enumerate}
\item   R. Benzi, G. Parisi, A. Sutera and A. Vulpiani; {\em Tellus} {\bf 34},
	      10 (1982); SIAM {\em J. Appl. Math} {\bf 43}, 565 (1983).
\item    {\em Noise in Nonlinear Dynamical Systems}; eds. F. Moss and
	     P.V.E. McClintock (Campridge Univ. Press, Cambridge 1989)
\item    F. Moss; {\em Stochastic resonance: from the ice ages
	     to the monkey's ear} in: Some Problems in Statistical
		  Physics, ed. G. Weiss, Frontiers in Applied Mathematics
		  (SIAM, Philadelphia 1992)
\item    P. Jung; {\em Phys. Rep.} {\bf 234}, 175 (1993)
\item    L. Gammaitoni, F. Marchesoni, E. Menichella-Saetta and
	       S. Santucci; {\em Phys. Rev. Lett.} {\bf 62}, 349 (1989)
\item    T. Zhou and F. Moss; {\em Phys. Rev. A} {\bf 41}, 4255 (1990)
\item    Proceedings on Nonlinear Dynamics in Optical Systems; eds.
	      M.B. Abraham, E.M. Garmire and P. Mandel (Optical Society of
		  America, Washington DC, 1991), Vol VII.
\item    L. Gammaitoni, M. Martinelli, L. Pardi and S. Santucci;
	      {\em Phys. Rev. Lett.} {\bf 67}, 1799 (1991)
\item    L. Gammaitoni, F. Marchesoni, M. Martinelli, L. Pardi
	       and S. Santucci; {\em Phys. Lett. A} {\bf 158}, 449 (1991);
\item   J. Heagy and W.L. Ditto; - preprint (1993), to be published in
	       {\em J. Nonlin. Sci.}
\item   P. Jung, U. Behn, E. Pantazelou and F. Moss; {\em Phys. Rev. A}
	{\bf 46}, R1709 (1992)
\item   A.R. Bulsara and G. Schmera; {\em Phys. Rev. E} {\bf 47}, 3734
	(1993)
\item   B.M. McCoy and T.T. Wu; {\em The Two-dimensional Ising Model}
	       (Harvard Univ. Press, Cambridge 1973)
\item  M. Acharyya and B.K. Chakrabarti; in {\em Annual Reviews of
       Computational Physics I.},
	 edited by D. Stauffer (World Scientific, 1994)
\item   N. Metropolis, A.W. Rosenbluth, M.N. Rosenbluth, A.H. Teller and
	       E. Teller; {\em J. Chem. Phys.} {\bf 21}, 1087 (1953)
\item   M.I. Dykman, D.G. Luchinsky, R. Mannella, P.V.E. McClintock,
	N.D. Stein and N.G. Stocks; {\em J. Stat. Phys.} {\bf 70}, 479 (1993)
\item   M.I. Dykman, R. Mannella, P.V.E. McClintock, N.D. Stein and
	N.G. Stocks; {\em J. Stat. Phys.} {\bf 70}, 463 (1993)

\end{enumerate}

\newpage
\begin{center}
\section*{Figure Captions}
\end{center}
\vspace{.30in}

$\:$ \\
\vspace{.15in}

{\bf Fig. 1.} Characteristic peak for the SR phenomenon in the plot of
the correlation ($\sigma=<B(t)\cdot M(t)>$) versus the systems temperature
($T_c=100$, $A=10$, $P=50$ and $N=5$).
\vspace{.5in}

{\bf Fig. 2.} Shape of the peak for three different values of the
magnetic fields amplitude ($A=5,10$ and $20$) ($T_c=100$, $P=50$ and $N=5$).
\vspace{.5in}

{\bf Fig. 3.}  Dependence of the resonance temperature versus the lattice
linear size for three different values of the period ($20$, $50$ and $200$)
($T_c=100$ and $A=10$).
\vspace{.5in}

{\bf Fig. 4.}  Variation of the resonance temperature ($T_r-T_c$) for
big lattices ($200\times 200$) against the period of the magnetic field.
The best-fit curve indicate: $T_r-T_c=557.264\cdot P^{-0.6335}$.

\end{document}